# Advanced Photodetectors for Hyperspectroscopy and Other Applications

I. Rodionov, J-M. Bidault, I. Crotty, P. Fonte, F. Galy, V. Peskov, O. Zanette

**Abstract-- Hyperspectroscopy is a new method of surface image taking, providing simultaneously high position and spectral resolutions which allow one to make some conclusions about chemical compositions of the surfaces. We are now studying applications of the hyperspctroscopic technique to be used for medicine. This may allow one to develop early diagnostics of some illnesses, as for example, skin cancer. For image taking advanced MCPs are currently used, sensitive in the spectral interval of 450-850 nm. One of the aims of this work is to extend the hyperspectrocpic method to the UV region of spectra: 185-280 nm. For this we have developed and successfully tested innovative 1D and 2D UV sealed photosensitive gaseous detectors with resistive electrodes. These detectors are superior MCPs due to the very low rate of noise pulses and thus due to the high signal to noise ratio. Other important features of these detectors are that they have excellent position resolutions - 30 µm in digital form, are vibration stable and are spark protected. The first results from the application of these detectors for spectroscopy, hyperspectroscopy and the flame detection are presented.**

## I. INTRODUCTION

There are several applications which required the detection of the weak Ultra Violet (UV) radiation with a high position resolution. As examples we will mention here only two recently appeared applications: hypespectroscopy and flame detection.

Hyperspectrocpy is a new method of surface image taking with simultaniosly high position and spectral resolutions (below 1nm) This gives enormous recognition power since such a method provides much more information than a human eye or usual colored pictures, for example it allows to make some conclusions about chemical compositions of surfaces. Thus the main function of hyperspectral remote sensing image data is to discriminate, classify, identify as well as quantify materials presented in the image. Another important application is subpixel object detection, which allows one to detect objects of interest with sizes smaller than the pixel resolution, and abundance estimation, which allows one to detect concentrations of different signature spectra present in pixels.

The Reagent Research Center developed and used this method for the analysis of earth surfaces from helicopters and satellites for environmental purposes such as for the search of spills from oils pipes [1]. We are now studying applications of the hyperspctroscopic technique to be used for medicine. This may allow one to develop early diagnostics of some illnesses, as for example, skin cancer.

The key elements in hyperspectroscopy are imaging photodetectors placed in the focal plane of the specially designed spectrographs. The main requirements to these detectors are: high sensitivity, high position resolutions (below 100 µm), high signal to noise ratio, low power consumption and of course, they should be stable in vibration.

For this application Reagent has developed advanced photodetectors based on a double step cascaded Micro Channel Plates (MCPs) combined with photocathodes (an active area of 3 $cm^2$) sensitive from 450 to 900 nm (see[1-4] and Fig. 1). Compared to commercially available MCPs combined with photocathodes (for example the Hamamatsu one) the Reagents's MCPs have several new design features allowing one to achieve a high photoelectron collection efficiency (with a broad collection efficiency plateau) and long-term stability at overall gains of $10^6$ and position resolutions of 12 µm (with a special coded multianode collector [2]). The frequency of the single electron noise pulses was between 10-100 kHz. A data acquisition system can handle counting rates of up to $10^6$Hz. For the on board data processing a compact supercomputer was developed by the "Reagent". An example of hyperspectroscopic obtained with these MCPs images is presented in Fig 2. One can clearly see from this figure the advantages of the hyperspectrocsopy over the usual image taking.

I. Rodionov is with the Reagent Research Center, Moscow, Russia, e-mail: i.rodionov@reagent-rdc.ru

J-M. Bidault is with the School of Mechanics at the Leonardo de Vinci University, 92916 Paris la Defence Cedex, France: Jean-Marc.BIDAULT@devinci.fr

I. Crotty is with the PH department of CERN, Geneva, Switzerland, e-mail: Ian.Crotty@cern.ch

P. Fonte is with the Ecole Superior de Ingenirea, Coimbra, Portugal, e-mail: fonte@lipc.fis.uc.pt

P. Caly is with the Research Division at the Leonardo de Vinci University, 92916 Paris la Defence Cedex, France, and CERN, Geneva, CH-1211 Switzerland, e-mail: Philippe.Galy@devinci.fr

V. Peskov is with the Research Division at the Leonardo de Vinci University, 92916 Paris la Defence Cedex, France, e-mail: vladimir.peskov@cern.ch

O. Zanette is with the School of Mechanics at the Leonardo de Vinci University, 92916 Paris la Defence Cedex, France e-mail: Olivier.ZANETTE@devinci.fr

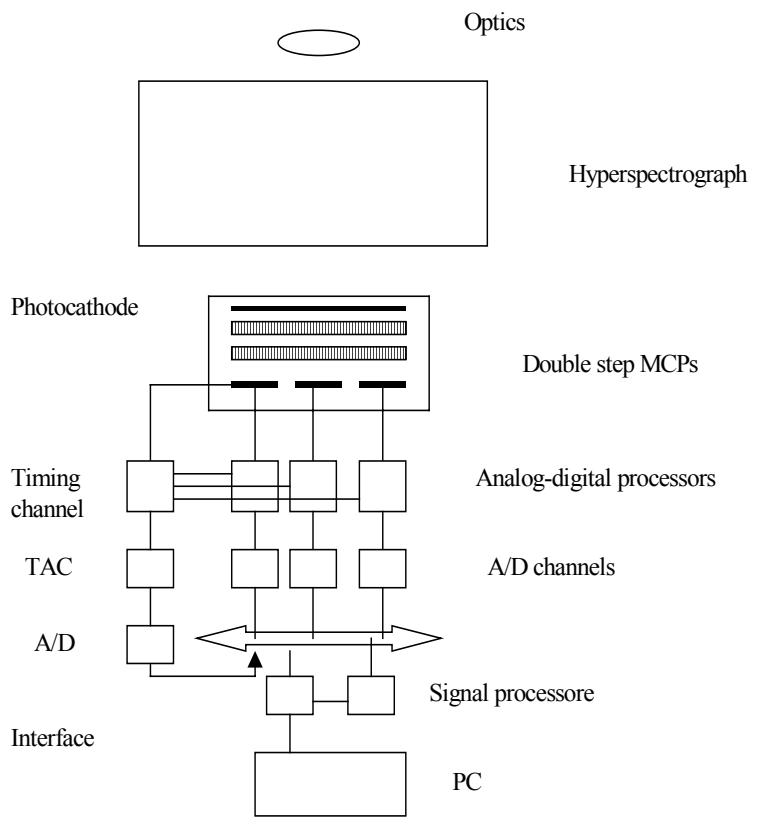

Fig1. A shematic drawing of the Reagent multianode MCP and the acquisition system

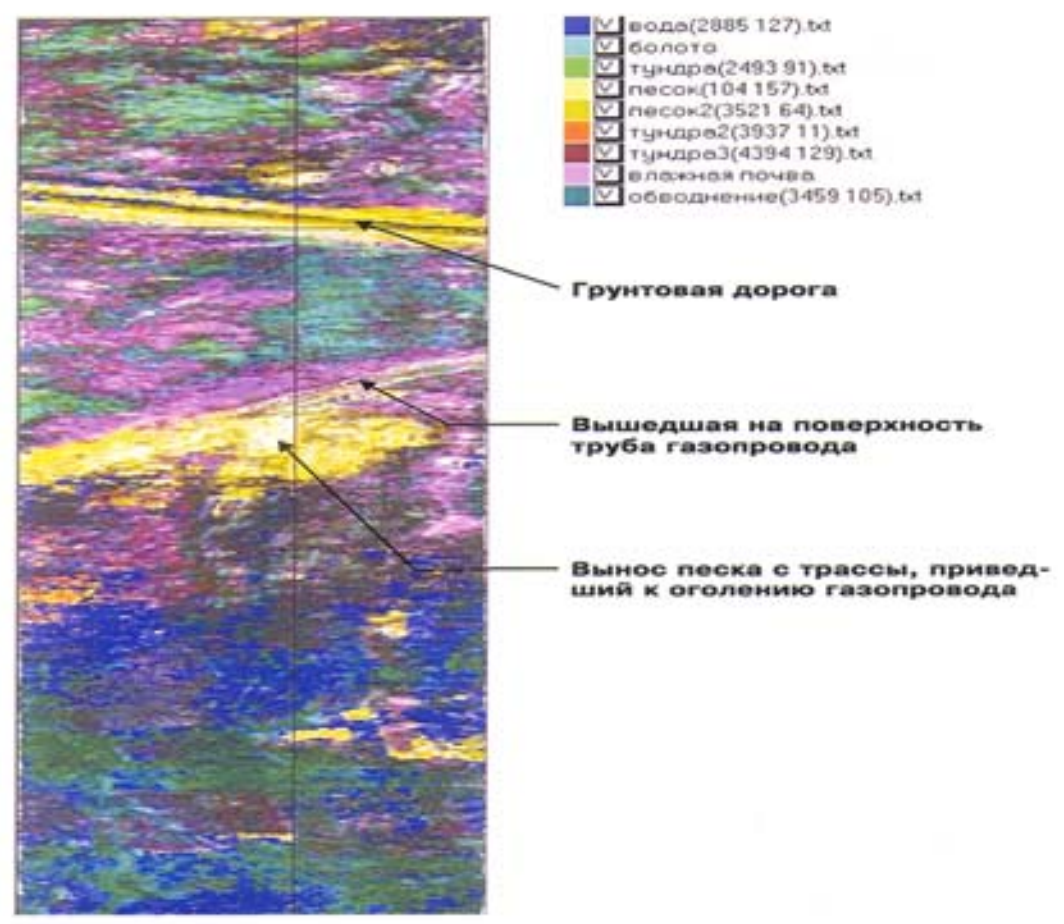

Fig2 A hyperspectroscopic computer synthesized imge of the earth surface [1]. One can identify the surface compositin, e.g. the yellow colour is sand, the rose colour is wheat earth, blue is water and the lighyt blue represets swam.

It will be very desirable to extend the hyperspectrocpic method to the UV region of spectra: 185-280 nm. This region is very attractive for the hyperspectroscpic measurements because the sunlight with a wavelength of 185-280 nm is blocked due to the absorption in the upper layers of the atmosphere (ozone) , but the atmosphere is transparent for this radiation on the ground level. This offers new technical possibilities in achieving materials and pattern recognitions; for example, it allows one to make unique hyperspectroscopic measurements using artificial UV light sources installed on the flying helicopter. The main problem in a such an approach is that the extremely low intensity of the scattered and reflected UV light should be detected on the strong background of the long wavelengths (with wavelength > 280 nm), produced by the sunlight. Existing MCPs are too noisy for these measurements [1].

The other new application which appeared just a few years ago is the flame visualization in the UV region of spectra. The main motivation comes from homeland security [5] and the possibilities to detect small forest fires or corona discharges on large distances [6,7]. As it was explained above, at the wavelength range of 260-280 nm daytime detection of UV signals emitted from the fire or the electrical discharges occurs with practically no background of solar radiation. Today the existing UV imaging detectors are mostly MCPs-based image intensifiers combined with narrow band filters transparent in the wavelength interval of 260-280 nm [8]. The sensitivity of these types of detectors is limited by the low transmission of the filters and high thermoelectronic noise level (typically $10^3$Hz per $cm^2$ of the photocathode surface).

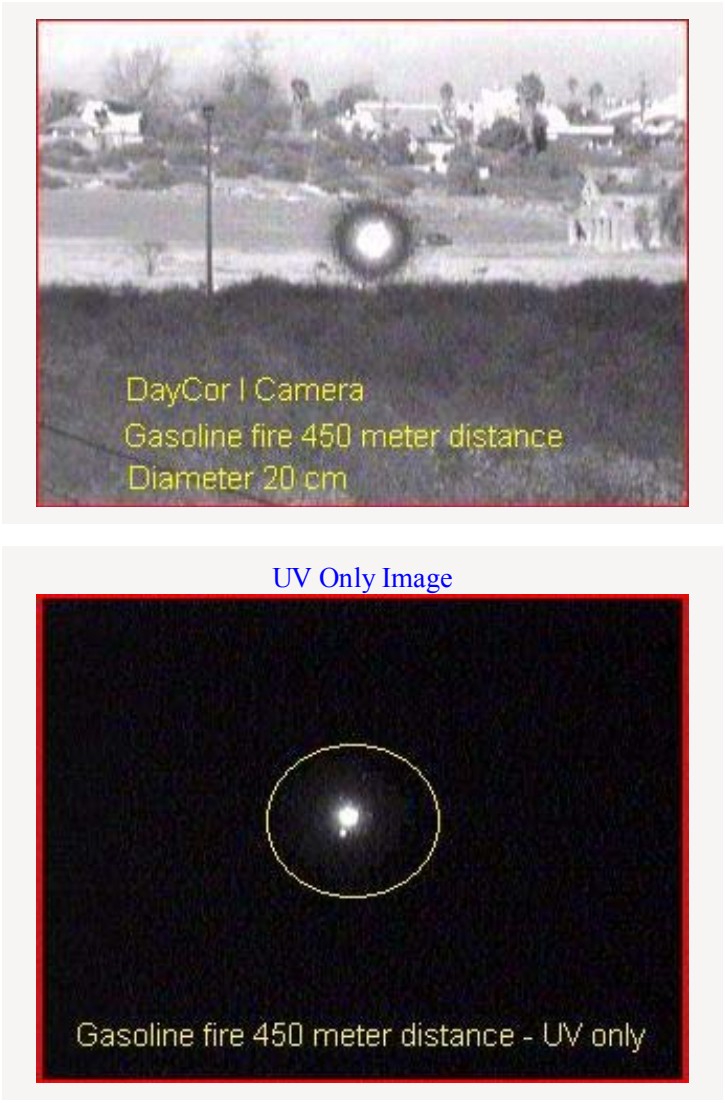

Fig. 3. Images of a small fame in the visisble and in the UV region of spectra , measured by DayCo camera [7].

For RICH applications we have recently developed flushed by gas RPCs combined with CsI photocathodes [9]. Their unique features are: high gas gains (they are spark protected), very good position resolutions (30-50μm) and extremely low thermoelectronic noise (<1 Hz per 1 $cm^2$ of the photocathode surface).

The aim of this work is to develop a new generation of photosensitive gaseous detectors with resistive electrodes oriented not only for RICH, but on wider range of applications such as hyperspectroscopy, flame detections and others. This requires the development of sealed version of these detectors as well as detectors combined with CsTe photocathode. In this paper we report our first results in this direction.

## II. Sealed Photosensitive RPCs and Their Application for Hyperspectrocopy

### A. *Experimental Set up*

The schematic drawing of the experimental set up is shown in Fig. 4. It contains a light source being emitted in the visible and in the UV interval of spectra, an object the surface of which should be investigated by measuring the spectra of its reflective light, a hypespectrograph and a detector of light: a PMT (EMI 9426 with a $MgF_2$ window) or photosensitive RPC installed in focal plane of the hyperspectrograph.

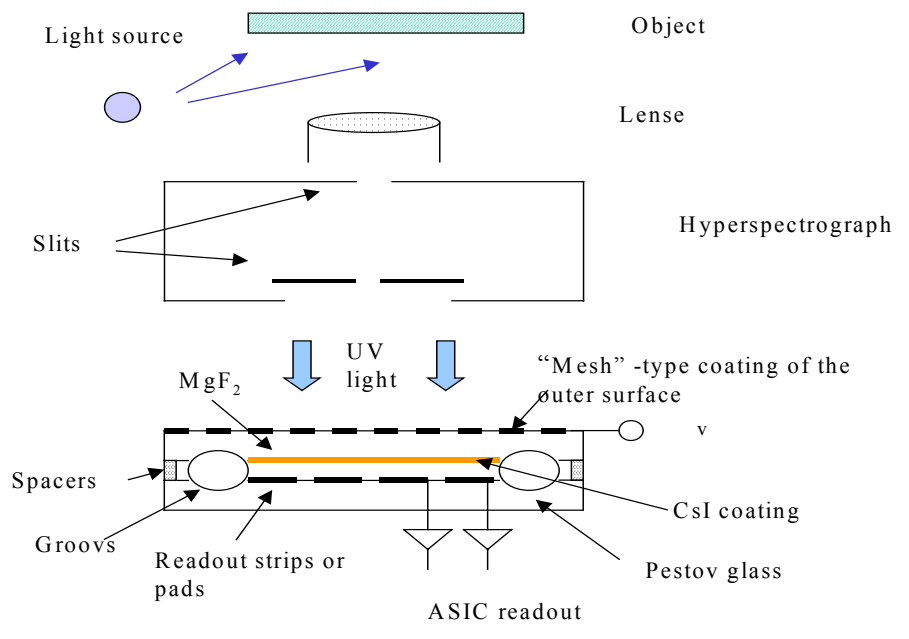

Fig.4. A schematic drawing of the experimental set up oriented on the spectroscopic and hypersepctrsocopic measurements in the UV region of spectra

The hyperspectrograph could work in two "modes": as a usual spectrograph with the installed an exit slit and as a hypespectrograph, when the exit slit was removed. In the first case the PMT was used to measure the spectra of the object. In the second case a 2D image was created on the entrance window of the photodetector: in one coordinate (y) it was an ID image of the object and in another coordinate (x) its spectra (see [1,10]). Thus in the case of a 2D detector one could measure at the same time the image of the object with a rather high position resolution and its spectra at each point of the object's surface.

The detailed description of the photosensitive RPC flushed by gas one can find in [9,10]. In this work for the first time we have developed and used the version of the RPC able to operate in a sealed gas envelope. This required a special preparation procedure: heating up to 50°C and pumping for one week, rinsing several times with a gas before a final filling with the gas mixture. Because it was quite difficult to build a sealed RPC with a 2D readout, for simplicity we used in this work RPCs with 1D readout only. First the readout strips of the detector were oriented along the y axis (spectral measurements) and then we rotated the detector on 90° and measured the ID image of the object at the selected wavelength.

### B. *Results*

Several various tests were done in which hyperspectroscopic images were taken with the photosensitive RPC and for comparison with MCPs as well. As an example we will present here results only with one of the tests performed. Fig. 5 shows a photograph of the two yellow stickers used as tests objects: a "dark" yellow (upper sticker) and a "light" yellow (lower sticker).

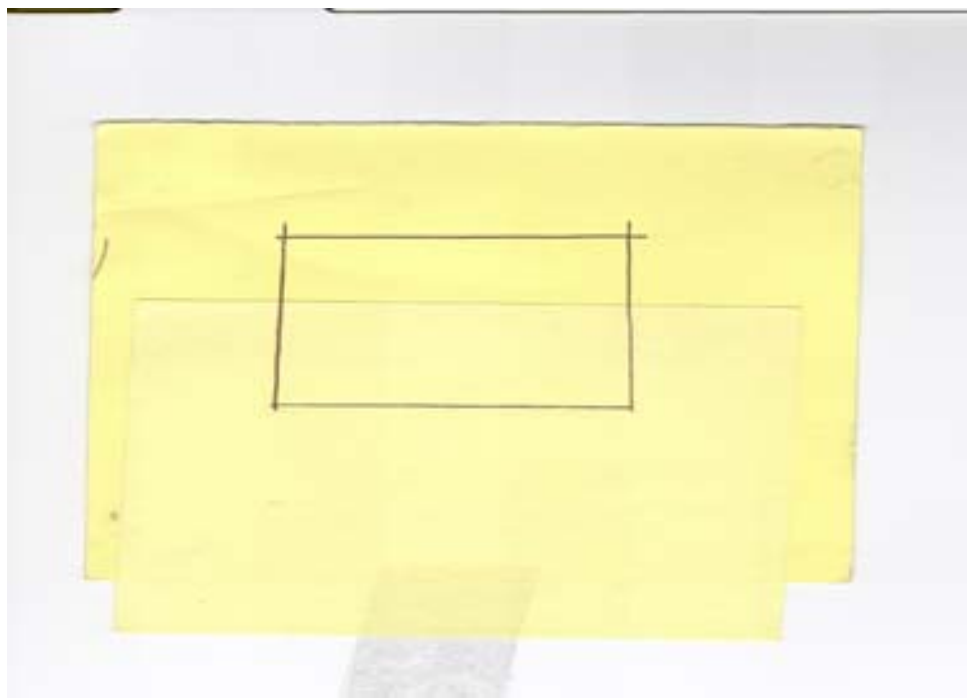

Fig.5. A photograph of the two stickers having a very close colour: "dark yellow (upper part) and" light" yellow (lower part)

In fact, they are hardy distinguished by eyes. However, by measuring the sticker's spectra one could easily distinguish between them-see Fig. 6 (the error of the measurements was less than the size of symbols presented in the figure). Note that below 250 nm signal to noise ratio of the PMT used was too low to reliably detect the weak reflective light from the stickers. In this spectral interval we used RPCs which had practically zero noise rates. As was explained above, the hyperspectrographic measurements we made were in two steps: first we performed spectral measurements and then turned the detector 90° and performed space resolved measurements at the given spectral interval. Examples of the obtained results are presented in Figures 7 and 8. From Fig 7. one can see that due to the very low rate of the noise pulses from our RPC we were able to detect the spectra below 200 nm. Fig. 8 shows the image of the boarder between the stickers measured at 194 nm.

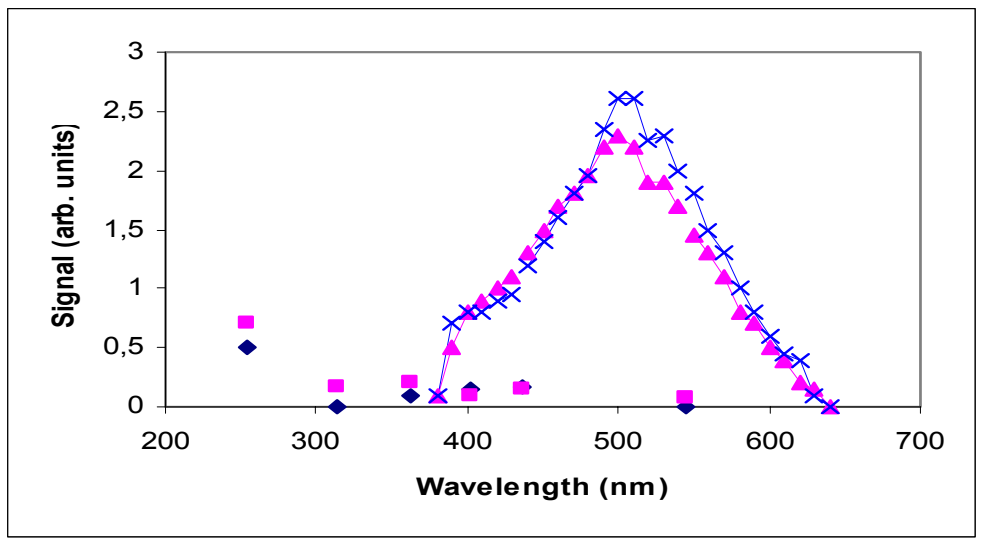

Fig.6. Spectra of the yellow stickers measure by a PMT. Rose squares - spectra of a "dark" sticker measured at wavelength corresponding to the strong lines from an Hg lamp, blue rhombus- the same measurements for a "light" sticker. Rose triangles – the reflective spectra of the "dark "sticker irradiated by a filament lamp, blue crosses- the same spectra measured for the "light" sticker

a)

b)

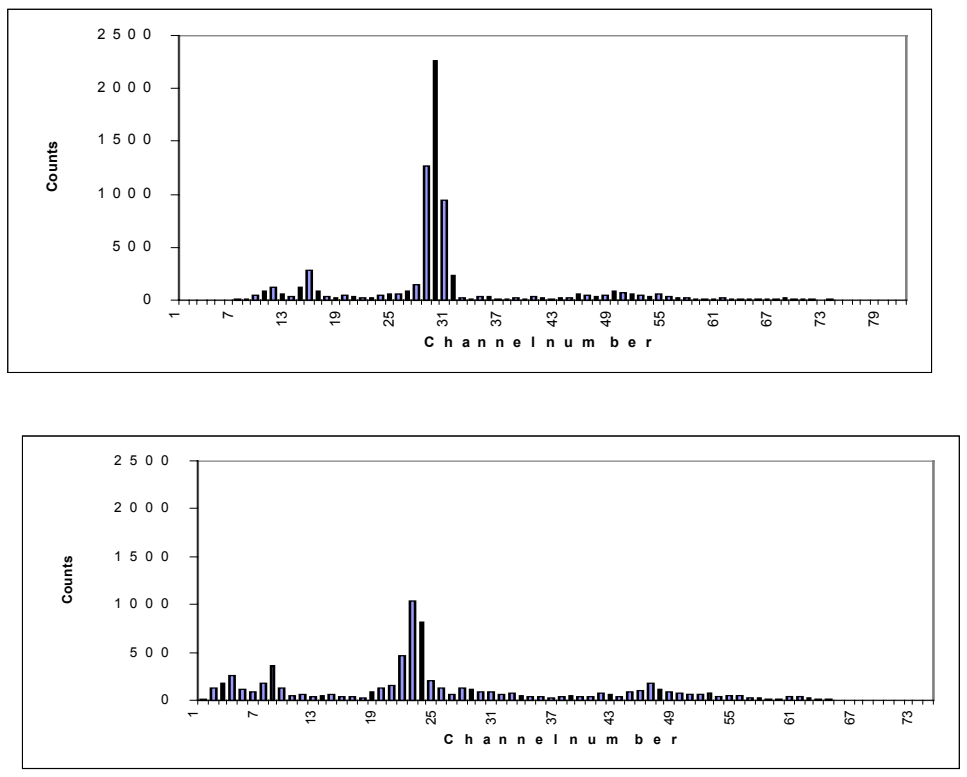

Fig. 7. On line digital image of the spectra of the yellow stickers (a-“light” and b-“ dark”) measured with photosensitive RPC

Thus we demonstrated that with the RPC we were able to detect a much weaker UV radiation than in the case of the PMT. For comparison we also performed some tests with the UV sensitive MCPs placed in the focal plane of the hyperspectrograph. The description of these results one can find in [1,10], we will just mention here that as in the case of the PMT, due to the high rate of the noise pulses from the MCP (>1000Hz), we were able to detect images only at wavelength >300 nm. Thus for applications requiring the detection of very weak UV radiations, photosensitive RPCs were superior to MCPs.

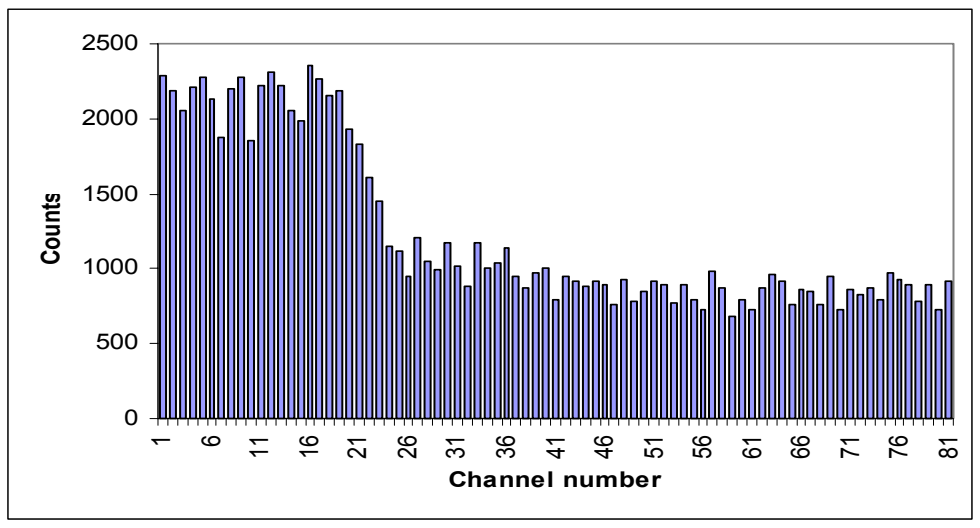

Fig. 8. On line digital image of a boarder between two yellow stickers measures with the RPC at 194 nm.

It was very important to demonstrate that the sealed RPC can operate stably for some time. Fig. 9 shows results of the RPC’s quantum efficiency (QE) monitoring during one month. One can see some degradation of the QE during the first week and then it remained quite stable. Note that in general, due its small thickness, the semitransparent CsI is much more sensitive to impurities in the gas than the reflective CsI photocathode, so we consider the obtained preliminary results as rather good. Better stability could be achieved with a gas chamber made of carefully chosen materials having low outgassing rate. The other possibility is to build an RPC combined with reflective CsI photocathode.

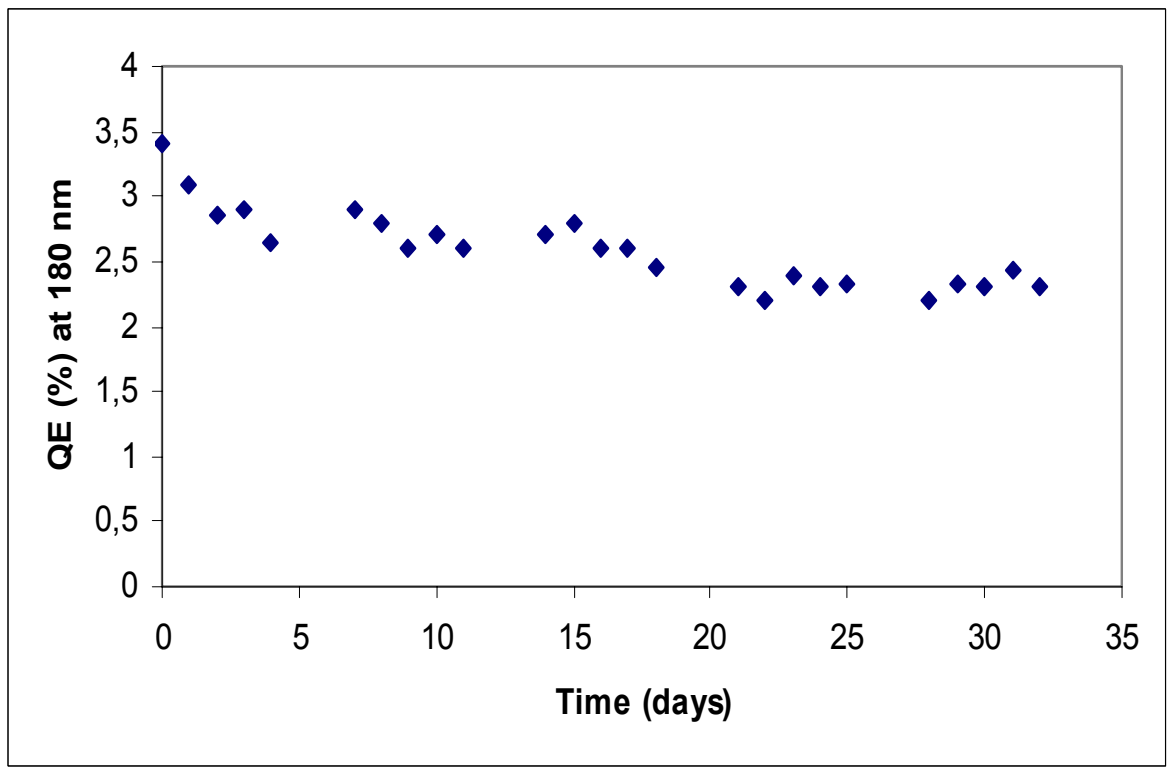

Fig. 9. Results of monitoring the RPCs QE with time

## III. The Application of The Photosensitive RPCs for Flame Detection

As was mentioned in the introduction, the other application where photosensitive RPCs could be used is in flame detection. One should note that MCP-based devices used today for this purpose have a QE a few times higher than our RPCs. However, as we already pointed out, the MCPs are quite noisy and are used in combination with narrow band filters which reduce their sensitivity at least on a factor of 10. This is probably why low noise RPCs may compete with MCPs in some measurements, especially in those which require the detection of a very weak UV radiation.

### A. *Experimental Set Up*

Our experimental set up is shown schematically in Fig. 10. It contains a test flame (a candle, for example), an RPC combined with a UV lens and an MCP- based UV image intensifier. In these particular measurements the RPC had a readout plate with metallic pixels (having pitches of 1,27 mm) similar to the one described in [11]. To simplify the readout electronics, the pixels were electrically connected in rows and amplifiers were connected to each row. This allowed one to perform 1D image measurements.

In some measurements, for example for the detection of smoke, several pulsed UV sources were used (pulsed $H_2$ lamps). These lamps gave well defined pulse signals at the readout elements. Appearance of the smoke on the line between the pulsed source and the RPC attenuated the UV light and caused a drop in the detected pulse amplitude which allowed one not only to detect smoke but also to measure its distribution in time and space.

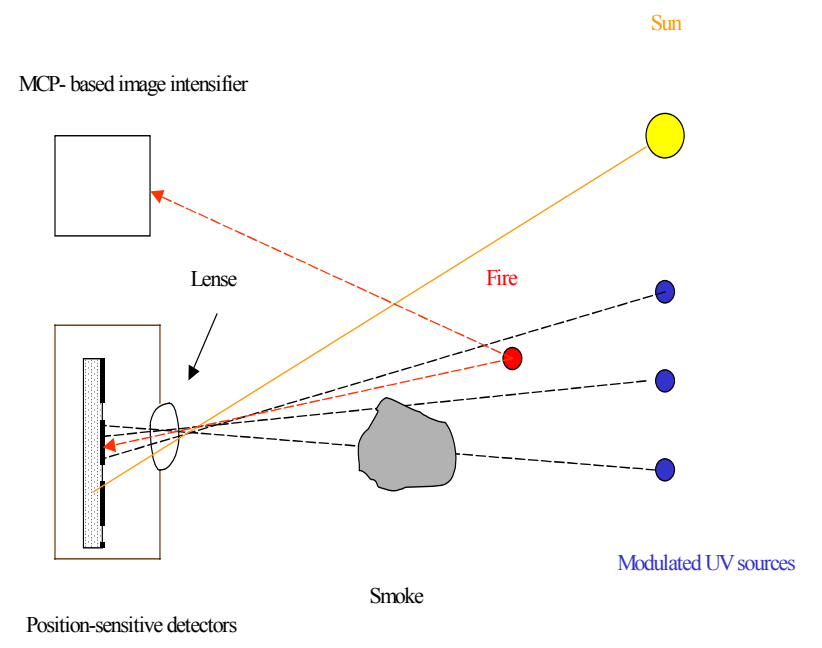

Fig.10. A schematic drawing of the experimental set up for visualization of the flames in the UV region of spectra and for detection of smoke

The performance and the sensitivity of the RPC was compared at various conditions with those of the MCP-based image intensifiers.

*B.. Results*

Obtained results allowed one to make quantative comparisons of the sensitivities of the RPC and the MCP. As an example in Fig. 11 one can see the image of the flame placed 1,5 m from the lens. The number of counts produced from the strips were measured during a period of 0,1 sec. The flame of the candle placed on a distance of 30 m from the RPC gave~$10^3$ Hz of the counting rate. Without the candle the counting rate from the same RPC operating in a fully illuminated room was <10 Hz. Note that 30 m was the maximum distance at which our image intensifier was able to detect the candle flame. So, for "in room" applications the sensitivity our RPC was comparable to the MCP. Unfortunately outside the building especially in direct sunlight, our RPC was too noisy compared to the MCP combined with a narrow-band filter. We are testing now another version of the photosensitive RPC able to operate in direct sunlight.
We also made some successful tests with smoke visualisation by using pulsed UV source. Some our earlier results in this direction one can find in [12].

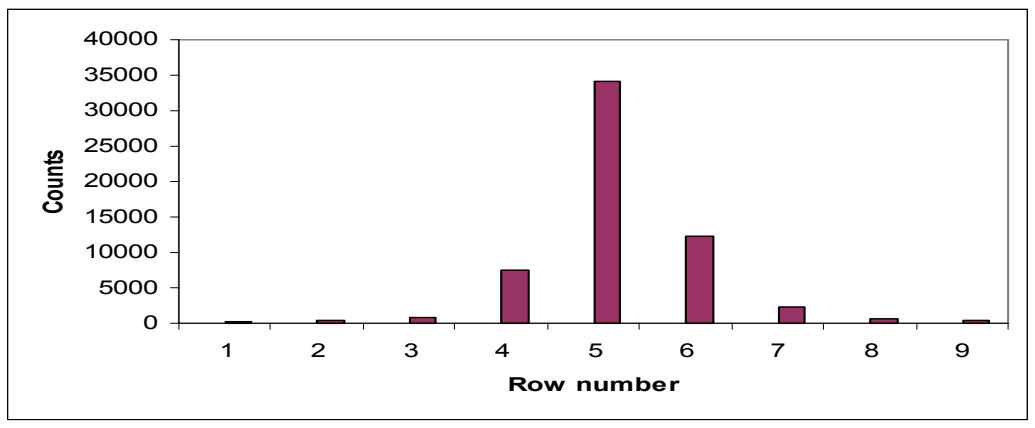

Fig. 11. A digital image of a flame from a candle obtained with the photosensitive RPC

The sensitivity of the RPC for the UV flame emission could be further improved if one succeeded to combine it with an CsTe photocathode, the QE curve of which is much better overlappes with the flame emission spectra than the QE curve of the CsI photocathode (see Fig. 12)

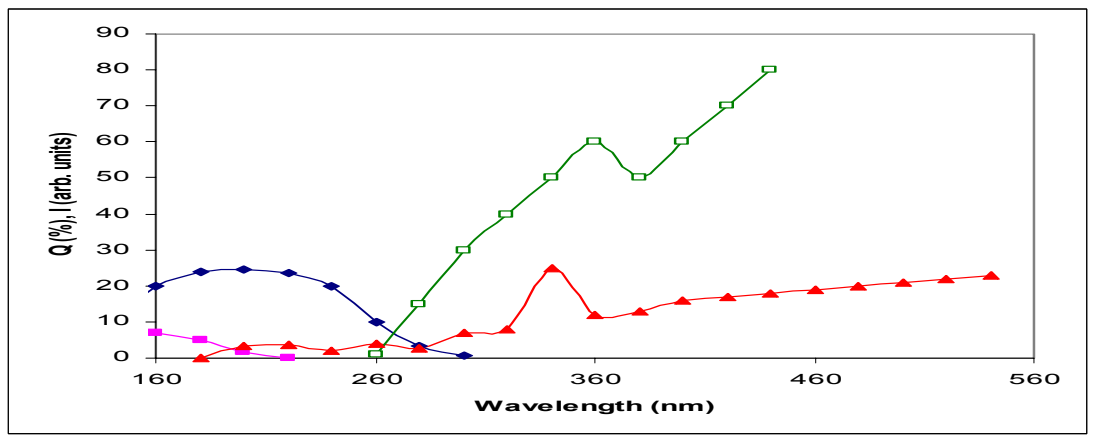

Fig. 12. The emission spectra of the flames in air (red) and the sun (green.) Rose the QE of the RPC combined with semitransparent CsI photocathode. Bleu- the QE of the CsTe photocathode [9]

The description of our afforts done in this direction is given in the next paragraph.

## IV. Development of Photosensitive Detectors with Resistive Electrodes combined with CsI and CsTe Photocathodes

Earlier we have demonstrated that gaseous detectors combined with CsTe photocathode can operate stably, but the maximum achievable gain was only ~ $10^3$ due to the feedback problems. In this work we have developed special devices: a hole- type structures with the resistive electrodes. We believe that such hole- type structures will allow one to suppress the feedback (and thus achieve higher gains) and resistive electrodes will made this detector spark protected. Below we describe our first tests of this innovative detector's design.

*A. Test of the gaseous detectors with the resistive electrodes CsI*

For simplicity the preliminary tests of this approach were done with G-10 hole- type structure developed earlier by our group [13] (see Fig. 13 and also [14]).

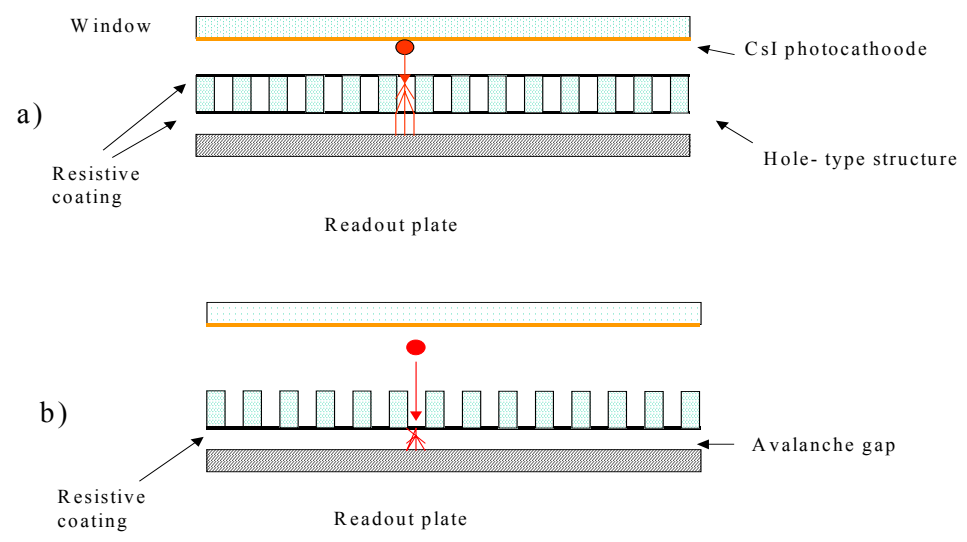

Fig. 13. A schematic drawing of the two designs of hole-type structures with resistive electrodes used in this work.

The Cu coating of the G-10 was almost fully removed and both surfaces were covered with graphite paint used in the usual RPCs. We named this structure a Resistive Electrode Hole Type Amplifier (REHTA). We have tested the operation of the REHTA in two modes: with the gas gain occurring inside the holes – "mode A" (see Fig. 13a) and with the gas gain occurring in the gap between the anode of the hole –type structure and the readout plate – "mode B" (see Fig. 13b). In first tests these detectors were combined with CsI photocathodes [15]. Some preliminary results are presented in Figs. 14 and 15. Fig. 14 shows gains vs. voltage for the Cu coated G-10 and for the REHTA operating in mode A (red symbols). As one can see due to the resistive coating REHTA allows higher gains to be achieved than with the Cu coated G-10 hole-type structure. We also tested the operation of the REHTA combined with CsI photocathodes. Usually hole- type detectors with metallic electrodes being combined with CsI photocathode operate at 5-10 less gains than one without the CsI photocathode [16]. However, in the case of REHTA the maximum achievable gain was only a factor of two less -see Fig. 14 (brown symbols). Fig. 15 shows gains vs. voltage for the RPC with the semitransparent CsI photocathode (described in the previous section) and for REHTA operating in B mode (see Fig. 13b). One can see that with REHTA gains higher than with the photosensitive RPC were achieved due to the feedback suppression by the hole type structure. These results encouraged us to make further tests with the CsTe photocathodes.

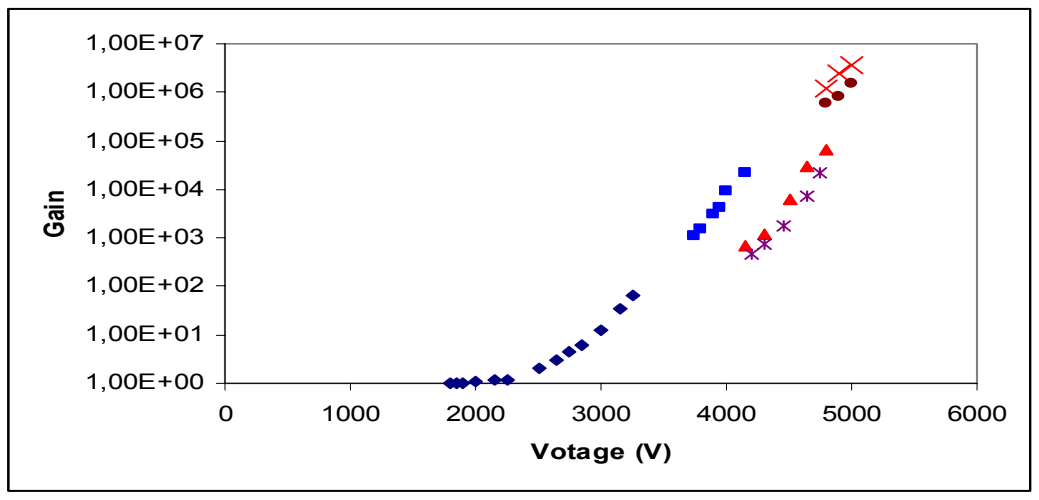

Fig. 14. Gain vs. voltage for the Cu coated G-10 hole type structure (blue) and for the REHTA operating in the mode A (red). Brown symbols represent the gain of the REHTA combined with an CsI photocathode. Gas mixture: Ar+5% isob =1 atm.

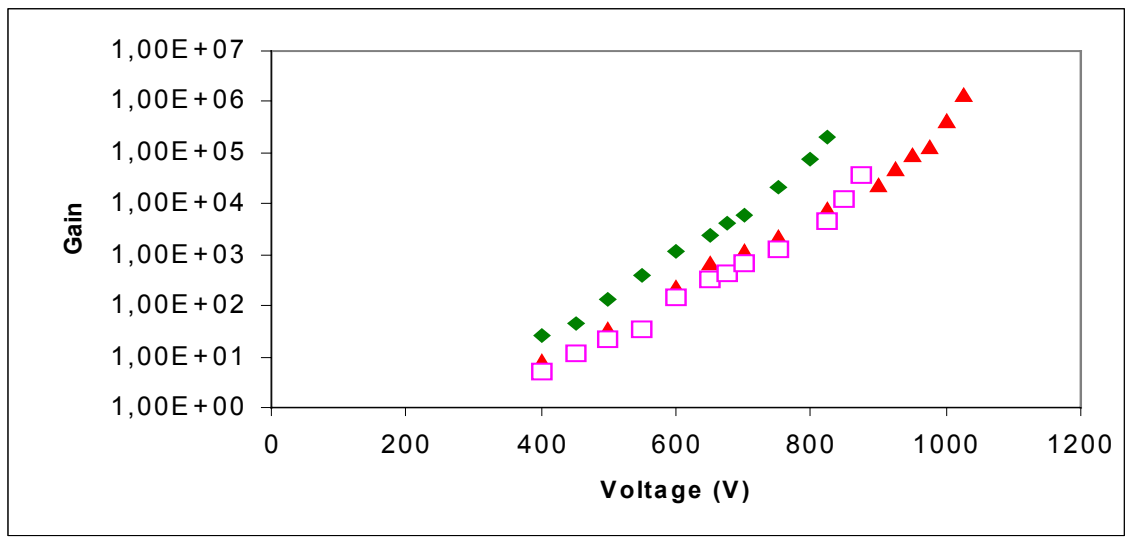

Fig. 15 Gain vs. voltage for RPCs and REHTA operating in mode B. Green rhombus – RPC combined with a semitranspatebt CsI photocathode, red triangles - REHTA combined with the semitransparent CsI photocathode, rose squares-REHTA combined with the CsTe/CsI photocathode. . All measurements were done in He+1%$CH_4$+EF gas mixture [9].

## *B. Development of the gaseous detectors combined with CsTe photocathodes*

The schematic drawing of the gas chamber for tests with the CsTe photocathodes is shown in Fig.16. The gas chamber had a $MgF_2$ window and manipulators allowing us to perform various mechanical movements with the detector's parts. A REHTA (operating in mode B) was installed inside this chamber together with a sealed (by In) cassette with an CsTe photocathode. Some tests were also done with an CsTe photocathode coated by 15-20 nm thick CsI protective layer (CsTe/CsI photocathode) designed to protect the CsTe photocathode from direct contact with the gas (to safe it from harming interactions with certain aggressive impurities in the gas). The technology of coating photocathodes by CsI protective layers was developed by us earlier [17] and latter in systematic works performed by Breskin group it was confirmed that CsI protective layer strongly improves the resistance of photocathodes to imprities in gases (see for [18] for more details). With the help of these manipulators the cassette containing the CsTe or CsTe/CsI photocathodes could be opened and combined with REHTA.

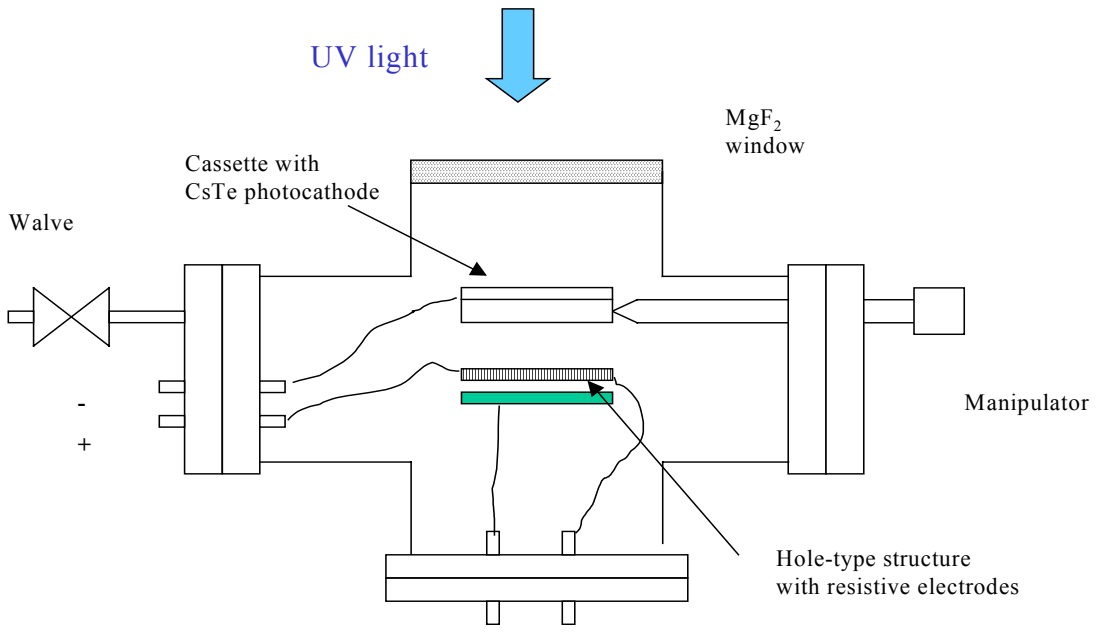

Fig. 16. A schematic drawing of the gas chamber for tests with CsTe photocathodes

Some preliminary results of our tests are presented in Figs. 17-18. Fig. 17 shows the QE of the CsTe and CsTe/CsI photocathodes at various conditions.

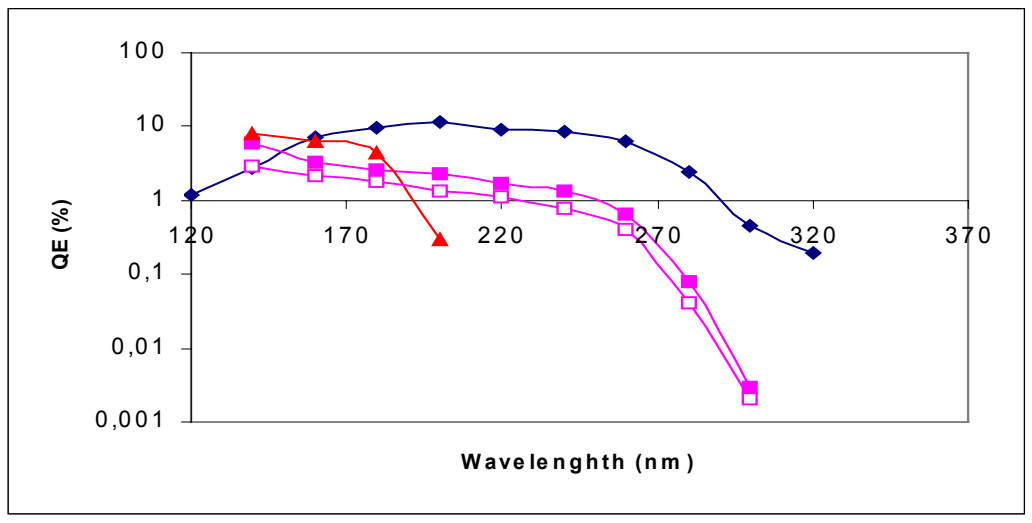

Fig. 17. QE of the photocathodes used in this work: bleu -QE of the CsTe photocathode measured in vacuum, rose filled squares QE of the CsTe photocathode in He+0,8%CH4+EF mixture [9], rose open squares-QE of the CsTe/CsI photocathode in He+0,8%CH4+EF; red triangles QE of the RPC combined with a CsI photocathode

From the data presented in Fig. 18 one can see that the CsTe photocathode could be very stable in clean gas conditions (without REHTA).

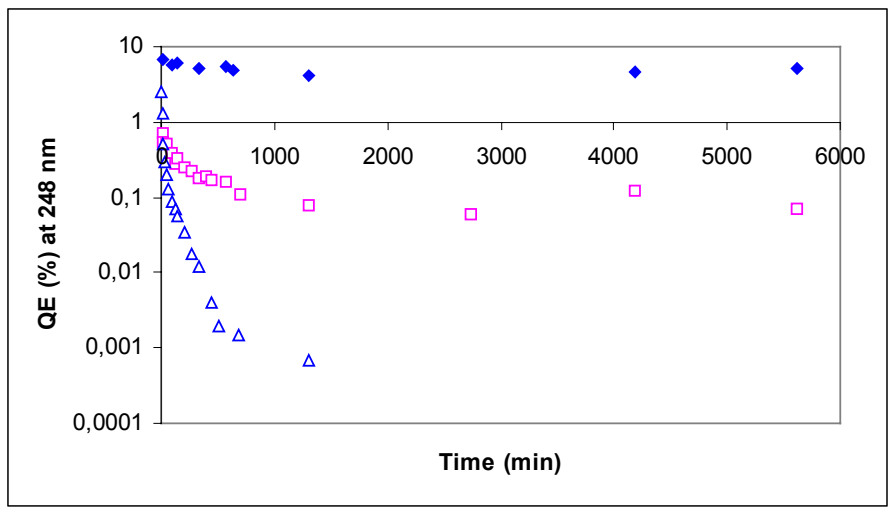

Fig. 18. QE vs. time for various photocathodes and at various conditions. Blue rhombus - QE of the CsTe photocathode installed in the gas chamber without a G-10 plate coated by graphite paint. Blue triangles- CsTe photocathode is combined with REHTA. Opern rose squares - the same goes for an CsTe photocathode coated by a CsI protective layer. All measurements were done in He+0,8%CH4+EF gas mixture [9].

However, in the presence of the REHTA, the CsTe photocathode degraded very rapidly due to the outgasing from the graphite paint. With REHTA some relative stability could be achieve only after two hours and only with the CsTe/CsI photocathode. In spite of big losses in the initial value of the QE, REHTA combined with such degraded photocathodes had sensitivity to the UV flame emission a few times higher than in the case of the CsI photocathode. These results indicate the perspectives of this approach. From Fig. 15 one can see that REHTA combined with CsTe/CsI photocathode can operate at rather high gains. The counting rate of the spurious pulsed form REHTA was rather low~10 Hz. This should be compared to the noise of CsTe photocathode in vacuum devices: for example, in the case of the PMT the noise rate was about 100Hz. This is because that in contrast to the "fast" photoelectrons created by the UV, the slow thermoelectrons cannot easily penetrate the protective layer. Thus the CsI layer plays a double role: as a protective layer preventing a direct contact of the CsTe photocathode with the gas and as a filter for slow electrons allowing to suppress thermoelectron emission from the CsTe.

## V. Conclusions

We have demonstrated that due to the very low rates of noise pulses, RPCs with CsI photocathodes and REHTA combined with CsTe/CsI photocathodes can offer a cheap and simple alternative to MCPs in some applications, for example in UV hyperspectroscopy or for "in room" flame detections. It is a very important that sealed gaseous detectors were tested in this work: only sealed detectors could be used in such applications as hyperspectroscopy or flame detection. There a lot of rooms for further improvements the characteristics of our devices. For example one can try to develop and use resistive coating with very low outgasing rates and thus achieve a better stability.

The question may arise: is it realistic to organise an industrial production of sealed photosensitive gaseous detectors at competitive prise? We think, yes. Several photonic companies already produce and sell at low price sealed gaseous detectors (for example UV flame detector Hamamatsu R2868). To day even several small research groups can manufacture sealed gaseous detectors combined with high efficiency photocathodes [19-21]. Their main advantage is the possibility to make them with large sensitive area. They also practically insensitive to magnetic fields, which make them attractive for some experiments. Thus we believe that sealed gaseous detectors may have a great future.